\documentclass[aps,prb,twocolumn,showpacs,preprintnumbers,amsmath,amssymb,superscriptaddress]{revtex4}%

\usepackage{graphicx}%
\usepackage{dcolumn}
\usepackage{amsmath}

\makeatletter
\def\btt#1{\texttt{\@backslashchar#1}}%
\DeclareRobustCommand\bblash{\btt{\@backslashchar}}%
\makeatother

\begin{document}

\preprint{PREPRINT (\today)}

\title{Oxygen isotope effect on structural parameters and on the thermal motion in La$_2$CuO$_4$ }

\author{Petra S.~H\"afliger}
\altaffiliation{Laboratory for Solid State Physics, ETH Zurich, CH-8093 Zurich, Switzerland}
\email{haeflig@phys.ethz.ch}

\author{Simon Gerber} \author{Ralph Chati}
\affiliation{Laboratory for Solid State Physics, ETH Zurich, CH-8039 Zurich, Switzerland}
%
\author{Vladimir Pomjakushin}
\affiliation{Laboratory for Neutron Scattering, ETH Zurich \& PSI, CH-5232 Villigen PSI, Switzerland}
\author{Kazimierz Conder}
\affiliation{Laboratory for Developments and Methods, ETH Zurich \& PSI, CH-5232 Villigen PSI, Switzerland}
\author{Ekaterina Pomjakushina}
\affiliation{Laboratory for Developments and Methods, ETH Zurich \& PSI, CH-5232 Villigen PSI, Switzerland}
\affiliation{Laboratory for Neutron Scattering, ETH Zurich \& PSI, CH-5232 Villigen PSI, Switzerland}
\author{Loic Le~Dreau}
\affiliation{Laboratory for Developments and Methods, ETH Zurich \& PSI, CH-5232 Villigen PSI, Switzerland}
\affiliation{Laboratory of Soft Chemistry and Reactivity, University of Rennes 1 UMR 6226, 35042 Rennes, France}

\author{Niels~B. Christensen}
\affiliation{Laboratory for Neutron Scattering, ETH Zurich \& PSI, CH-5232 Villigen PSI, Switzerland}

\author{Henrik~M. R{\o}nnow}
\affiliation{Laboratory for Quantum Magnetism, EPF Lausanne, CH-1015 Lausanne, Switzerland}

\begin{abstract}
We studied the oxygen isotope effect (OIE) on the zero point motion and the thermal motion as well as on structural parameters in  La$_2$CuO$_4$ by means of  high-resolution neutron diffraction experiments. We found a negative OIE  on the lattice parameters (-0.01\%).  The OIE on the isotropic thermal parameters turned out to be positive for lanthanum, copper and negative for the oxygen atoms, respectively. The Rietvield refinement of the  anisotropic thermal parameters of the different directions yielded the same sign of the OIE  for each ion. Our analysis shows that the OIE on  isotropic thermal parameters is most pronounced for La at 15~K (up to 100\%) which we found to be originating mainly from the thermal motion in $x$-direction as determined from the refinement of the anisotropic thermal parameters.
\end{abstract}


\maketitle
\section{Introduction}
Possibly the largest enigma in contemporary solid state physics is how doping holes into the antiferromagnetic planes can lead to high transition temperature superconductivity in the cuprates. Early after their discovery, Anderson proposed an explanation in terms of resonating valence bonds (RVB)~\cite{Anderson3} based on the fact that the parent compound exhibits long-range antiferromagnetic order~\cite{kastner}. Indeed quantum effects seem to play an essential role in the cuprates: For instance the observed zone boundary spin-wave dispersion  in La$_2$CuO$_4$~\cite{coldea} may be a fingerprint of an intrinsic quantum effect typical for a 2D quantum Heisenberg antiferromagnet~\cite{henrik}. The importance of quantum fluctuations should also be reflected in zero-point motion effects. In the case of C$_{60}$ the extent of zero-point motion effects turned out to be significant~\cite{kohanoff} in a sense that they may be responsible for a substantial renormalization of electron-phonon coupling. In the lanthanum cuprates, this point has not been addressed experimentally  so far. In particular no information about the oxygen isotope effect (OIE) on the zero point motion is available.

\noindent
Isotope effects on electronic properties in the cuprates are well known, the superconducting transition temperature decreases with oxygen isotope substitution~\cite{keller} whereas the pseudogap temperature raises as has been demonstrated e.g. by means of neutron cyrstal-field spectroscopy~\cite{petra}.
Regarding  isotope effects on magnetic properties only little is known so far.  Recently Khasanov {\it et al.}~\cite{rustem} performed a detailed muon-spin rotation and magnetization study of the isotope dependence of magnetic quantities, they found that the antiferromagnetic ordering and spin-glass ordering temperature in Y$_{1-x}$Pr$_x$Ba$_2$Cu$_3$O$_{7-\delta}$ exhibit a large OIE  in the regime where superconductivity and antiferromagnetic order coexist. A huge OIE on the spin glass temperature was also found in Mn-doped LSCO at low doping  while the isotope effect is small in Mn-free samples~\cite{shengelaya}. These unusual effects could arise from the isotope dependent mobility of the charge carriers~\cite{bussmann}.
As for the undoped compound La$_2$CuO$_4$, the N\'eel temperature $T_N$  slightly decreases upon oxygen isotope substitution~\cite{zhao}. These findings  were assumed to originate from structural changes upon oxygen isotope substitution ~\cite{Hanzawa}. But so far only the isotope effect on the lattice constants and the following orthorhombicity have been measured. Thus there is need for a careful neutron diffraction study to extract the OIE on structural parameters and on thermal motion which might be relevant for $J$, the coupling constant of the in-plane antiferromagnetic superexchange interaction between nearest-neighbor Cu$^{2+}$ spins.
\\  
To do so we perforemd a high-resolution neutron diffraction study. Neutron diffraction is an excellent tool to study thermal and zero point motion since the neutron cross section is directly proportional to the Debye-Waller factor.

\section{Experiments}
Polycrystalline samples of La$_2$CuO$_4$ were prepared using conventional solid state reaction. Oxygen isotope exchange was performed using the procedure described previously \cite{conder}. 
The oxygen content was determined by thermogravimetric hydrogen reduction~\cite{conder2}. Magnetization measurements were performed using a Quantum Design MPMS in fields ranging from 20~mT to 6~T at temperatures between 4 and 300~K on La$_2$Cu$^{16}$O$_4$  
and La$_2$Cu$^{18}$O$_4$.
\\
Subsequent neutron diffraction experiments were performed on the high-resolution diffractometer HRPT~\cite{hrpt} at SINQ~\cite{sinq} located at the Paul Scherrer Institute, Switzerland. The experiments were carried out at a wavelength $\lambda=1.1545$ \AA. The La$_2$Cu$^{16}$O$_4$  and La$_2$Cu$^{18}$O$_4$ sample were each contained in a 8mm-diameter Vanadium container which was then mounted into a closed cycle refrigerator  in order to reach temperatures between 15 K and 290 K. 
High statistics data were taken at 15 K and 290 K ($3.5\cdot 10^7$ counts), whereas points at temperatures $15<T<290$ K were obtained at intermediate statistics ($2.5\cdot 10^6$ counts).

\begin{figure}
 \centering\includegraphics[width=0.5\textwidth]{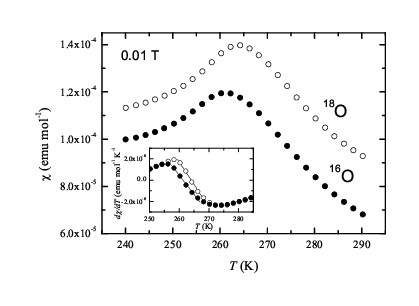}
 \caption{{Susceptibility vs. temperature for La$_2$Cu$^{16}$O$_4$ (solid circles) and La$_2$Cu$^{18}$O$_4$ (open circles). The inset shows the derivative of the susceptibility.}
\label{chi}}
\end{figure}

\begin{figure}
 \centering\includegraphics[width=0.5\textwidth]{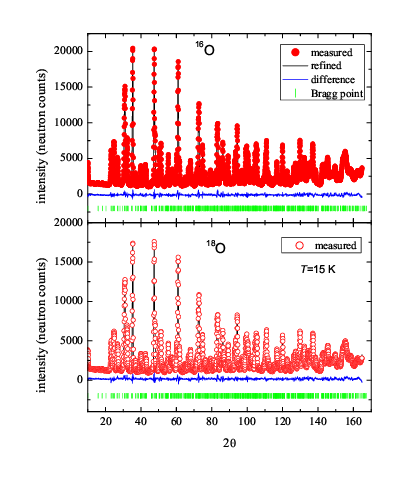}
 \caption{{Neutron diffraction data at $15~K$ for La$_2$Cu$^{16}$O$_4$ (upper panel) and La$_2$Cu$^{18}$O$_4$ (lower panel). The solid black line corresponds to a Rietvield refinement as explained in the text. The difference between measured  and calculated profile is shown below. Tick marks at the bottom represent the position of allowed Bragg reflections in the $Bmba$ space group.}
\label{nd}}
\end{figure}

\section{Results}
\subsection{Macroscopic measurements}
Susceptibility $\chi(T)$ versus temperature is shown in Figure~\ref{chi} in an applied field of 0.1~T for the $^{16}$O and the $^{18}$O sample.  A clear peak in the susceptibility is seen at $261.4\pm0.1$~K and $263.8\pm0.1$~K for the  $^{16}$O and the $^{18}$O sample, respectively. The derivative $d\chi/dT$ is shown in the inset of Fig.~\ref{chi}.

\subsection{Neutron diffraction}
\begin{figure}
 \centering\includegraphics[width=0.5\textwidth]{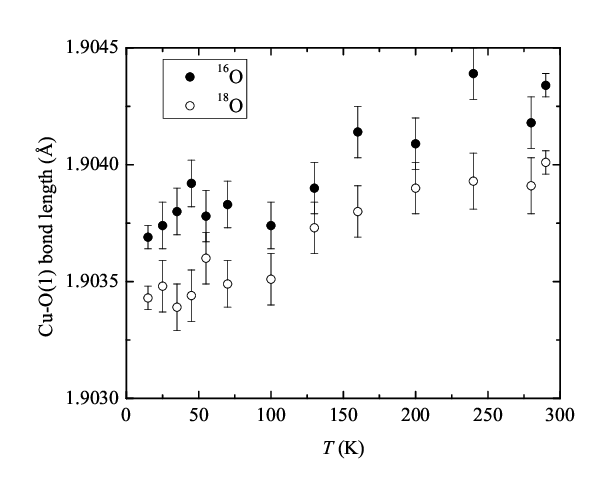}
 \caption{{Cu-O1 bond lengths vs. temperature $T$ for La$_2$Cu$^{16}$O$_4$ (solid circles) and La$_2$Cu$^{18}$O$_4$ (open circles).}
\label{cuo1}}
\end{figure}

The intensity pattern from neutron diffraction were refined using the orthorhombic nonstandard space group {\it Bmba} which is isomorphic to the standard space group {\it Cmca} (No.~64.), with atomic positions O1$=(1/4,1/4,z)$, O2$=(0,y,z)$, La$=(0,y,z)$ and Cu$=(0,0,0)$. The occupancy of all atoms was set to the full value, the oxygen content was fixed to 4.0. Isotropic and  anisotropic Debye-Waller temperature factors (short: thermal parameters) were refined for each data set.
The results for the lattice constants, atomic positions and Cu-O1 bond lengths obtained from a refinement with isotropic parameters are reported in Table~\ref{position}. Apparently there is no OIE on atomic positions within the resolution of the diffractometer and the fitting procedure. The results obtained using an anisotropic refinement are identical within the range of error. 
\\
In the following we express the OIE on a quantity $x$ such as lattice parameters $a$, $b$, $c$ etc. by $\Delta x= \frac{x^{18}-x^{16}}{x^{16}}$ ($x^{16}$and $x^{18}$ are short for the quantity $x$ measured in La$_2$Cu$^{16}$O$_4$ and La$_2$Cu$^{18}$O$_4$, respectively). The results for OIE on the lattice parameters and the Cu-O1 bondlength at 15~and 290~K are summarized in Table~\ref{lattice_oie}, the OIE turned out to be in the range of 0.01 \% for all the parameters. Interestingly the orthorhombicity is more pronounced in the $^{18}$O sample: we found that $a_{16}/b_{16}=0.98435(1)$ at $T=15$~K  and 0.99101(1) at $T=290$~K whereas $a_{18}/b_{18}=0.98419(1)$ at 15~K  and  and 0.99093(1) at 290~K. The Cu-O1 bondlength vs. temperature is displayed in Fig.~\ref{cuo1}.

\title{Broad table}
\begin{table*}
\begin{tabular}{c|ccc|ccccc|c} 
\hline\hline
 & $a$ (\AA) & $$b$ (\AA)$ & $c$ (\AA) & $y$(La)  & $z$(La) & $z$(O1) & $y$(O2) &  $z$(O2) & Cu-O1 (\AA) \\ \hline
15~K & & & & & & & & & \\
$^{16}$O & $5.33305(2)$ & $5.41783(3)$ & $13.10260(7)$ & $-0.00839(13)$ & $0.36149(4)$ & $-0.00832(6)$ & $0.04101(12)$ & $0.18302(8)$ & $1.90369(5)$\\
$^{18}$O & $5.33185(3)$ & $5.41751(3)$ & $13.09979(8)$ & $-0.00855(14)$ & $0.36158(4)$ & $-0.00834(6)$ & $0.04110(13)$ & $0.18293(8)$ & $1.90343(5)$\\ \hline
290~K & & & & & & & & & \\
$^{16}$O & $5.35479(3)$ & $5.40338(3)$ & $13.14810(9)$ & $-0.00683(19)$ & $0.36135(4)$ & $-0.00746(8)$ & $0.03479(17)$ & $0.18308(9)$ & $1.90434(5)$\\
$^{18}$O & $5.35385(3)$ & $5.40283(3)$ & $13.14614(9)$ & $-0.00655(20)$ & $0.36138(4)$ & $-0.00736(8)$ & $0.03463(17)$ & $0.18291(9)$ & $1.90401(5)$\\ \hline\hline
\end{tabular}
\caption{Lattice parameters, atomic coordinates and the Cu-O1 bond lengths for La$_2$Cu$^{16}$O$_4$ and La$_2$Cu$^{18}$O$_4$ at $15~$ and $~290~K$ as obtained in an isotropic refinement. In the nonstandard space group {\it Bmba} (isomorphic to the standard {\it Cmca}, No.~64) used in the present work, the atomic positions are La~$[0,y,z]$, Cu~$[0,0,0]$, O1~$[1/4,1/4,z]$, O2~$[0,y,z]$.} 
\label{position}
\end{table*}

\begin{table}
\begin{tabular}{c|ccc|c}
\hline\hline
$T$ & $\Delta a$ $[10^{-4}]$  & $\Delta b$ $[10^{-4}]$ & $\Delta c$ $[10^{-4}]$ & $\Delta$~Cu-O1 $[10^{-4}]$\\ \hline
15~K & $-2.25(9)$ & $-0.59(11)$ & $-2.14(14)$  & $-1.37(53)$\\
290~K & $-1.76(13)$ & $-1.02(11)$& $-1.37(53)$ &  $-1.73(53)$
\\ \hline\hline
\end{tabular}
\caption{OIE on the lattice parameters in La$_2$CuO$_4$ as obtained by an isotropic refinement.}
\label{lattice_oie}
\end{table}

\begin{figure}
 \centering\includegraphics[width=0.5\textwidth]{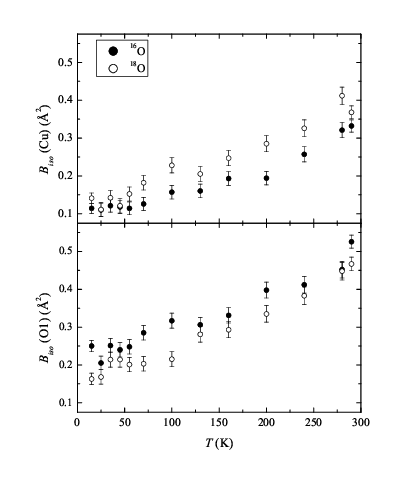}
 \caption{{Temperature dependence of the isotropic thermal parameters $B_{iso}$ for Cu and O1.}
\label{biso}}
\end{figure}

In Fig.~\ref{biso} the temperature evolution of the isotropic parameters $B_{iso}$ for Cu and O1 is illustrated for both the $^{16}$O and $^{18}$O sample. The $B_{iso}$ parameters increases with temperature, but below temperatures of approximately 100 K a saturation sets in. We note that the points at 15~K and at 290~K are measured with high statistics as opposed to the intermediate points which explains the larger error bars and the offset when going from the high-statistics to the low-statistics data. The results from the high-statistics data are summarized in Table~\ref{baniso}. The magnitude of the $B_{iso}$-parameters scale qualitatively with the inverse mass of the atoms, more precisely $B_{iso}(O2) >B_{iso}(O1)>B_{iso}(Cu)>B_{iso}(La)$ at 15~K and $B_{iso}(O2) >B_{iso}(O1)>B_{iso}(La)>B_{iso}(Cu)$ at 290~K, respectively. 
\\
The corresponding OIE on $B_{iso}$-parameters are given in Table~\ref{biso_oie}. We found that $B_{iso}$ of the oxygen atoms O1 and O2 exhibit a negative OIE, whereas the OIE is positive for La and Cu. 

\begin{table*}
\centering
\begin{tabular}{ll|cccc|cccc}
\hline\hline
 & & $15~K$ & & & & $290~K$ & & & \\ \hline
 &  & $B_{iso}$ & $U_{11}$ & $U_{22}$ & $U_{33}$ &   $B_{iso}$ & $U_{11}$ & $U_{22}$ & $U_{33}$\\ 
\hline
$^{16}$O & La & 0.075(6) &  0.00059(16) & 0.00171(18) & 0.00060(20)  & 0.367(7) & 0.00526(17) & 0.00553(21) & 0.00371(18) \\
$^{18}$O & &0.115(6) & 0.00113(18) & 0.00218(19) & 0.00111(32) &  0.407(7) & 0.00540(19) & 0.00612(22) & 0.00458(18)\\ \hline

$^{16}$O & Cu & 0.109(8) & -0.00019(20) & 0.00214(23) & 0.00212(35) &  0.320(10) & 0.00145(26) & 0.00386(28) & 0.00764(35) \\ 
$^{18}$O & & 0.144(9) & 0.00028(28) & 0.00263(30) & 0.00248(35) &  0.360(10) & 0.00192(18) & 0.00443(30) & 0.00808(35)\\ \hline

$^{16}$O & O1 & 0.234(9) & 0.00237(26) & 0.00330(25) & 0.00289(35)  & 0.520(11) & 0.00374(25) & 0.00480(27) & 0.01207(36)\\ 
$^{18}$O & & 0.183(9)  & 0.00176(27) & 0.00256(27) & 0.00235(35) &  0.471(11) & 0.00357(26) & 0.00430(28) & 0.01083(36)\\ \hline

$^{16}$O & O2 & 0.416(9) & 0.00693(24) & 0.00431(31) & 0.00389(26) & 0.970(13) & 0.01923(35) & 0.01276(44) & 0.00498(26)\\ 
$^{18}$O & & 0.339(10) & 0.00601(24) & 0.00324(31) & 0.00302(26) &  0.915(13) & 0.01862(36) & 0.01118(44) & 0.00456(35)\\ 
\hline\hline
\end{tabular}
\caption{B$_{iso}$-parameters in \AA$^2$ and $U_{ii}$-parameters ($1\leq i \leq 3$) in \AA$^2$ of  La$_2$CuO$^{16}$$_4$ and La$_2$Cu$^{18}$O$_4$ at $15~$and$~290~K$ for an isotropic and anisotropic refinement, respectively.}
\label{baniso}
\end{table*}

\begin{table}
\begin{tabular}{l|c|c}
\hline\hline
 & $15~K$ &  $290~K$  \\ \hline
 & $\Delta B_{iso}$ &    $\Delta B_{iso}$ \\ \hline
La & $0.53(16)$ &  $0.11(4)$  \\
Cu & $0.32(15)$ & $0.13(6)$ \\
O1 & $-0.22(8)$ &  $-0.094(4)$  \\
O2 & $-0.19(4)$ &  $-0.06(3)$
\\ \hline\hline
\end{tabular}
\caption{OIE on isotropic  thermal parameters $\Delta B_{iso}$ for  La$_2$CuO$_4$  at $15~$and$~290$~K.}
\label{biso_oie}
\end{table}

The results of the anisotropic thermal parameters  are also reported in Table~\ref{baniso}. Since data were collected from a powder sample it is only possible to refine the diagonal elements $U_{11}$, $U_{22}$  and $U_{33}$. 
From the order of magnitude of $U_{ii}$, $1\leq i\leq 3$ we deduce in which direction the movements are the largest: at 15 K $U_{max}=U_{22}$ for La, Cu and O1 and $U_{max}=U_{11}$ for O2. At 290 K we find $U_{max}=U_{22}$ for La, $U_{max}=U_{11}$  for O2 and  $U_{max}=U_{33}$ for Cu and O1.  Similarly to the isotropic refinement the oxygen atoms showed a negative OIE, whereas La and Cu atoms had a positive OIE in all the directions. In particular we found that the OIE on $U_{11}$ is huge for La ($\sim 90$ \%)and Cu ($\sim 47$ \%) at 15 K. 
\begin{figure}
 \centering\includegraphics[width=0.5\textwidth]{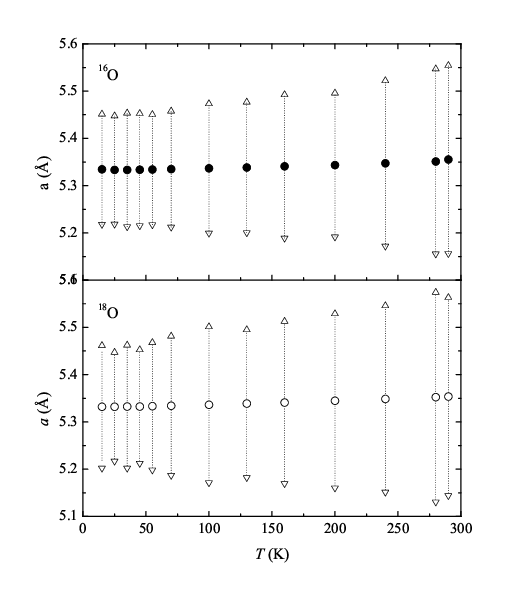}
 \caption{{Lattice parameter $a$ as a function of temperature with displacement due to thermal motion as deduced from Biso parameters.}
\label{lattice_biso}}
\end{figure}

\begin{figure}
 \centering\includegraphics[width=0.5\textwidth]{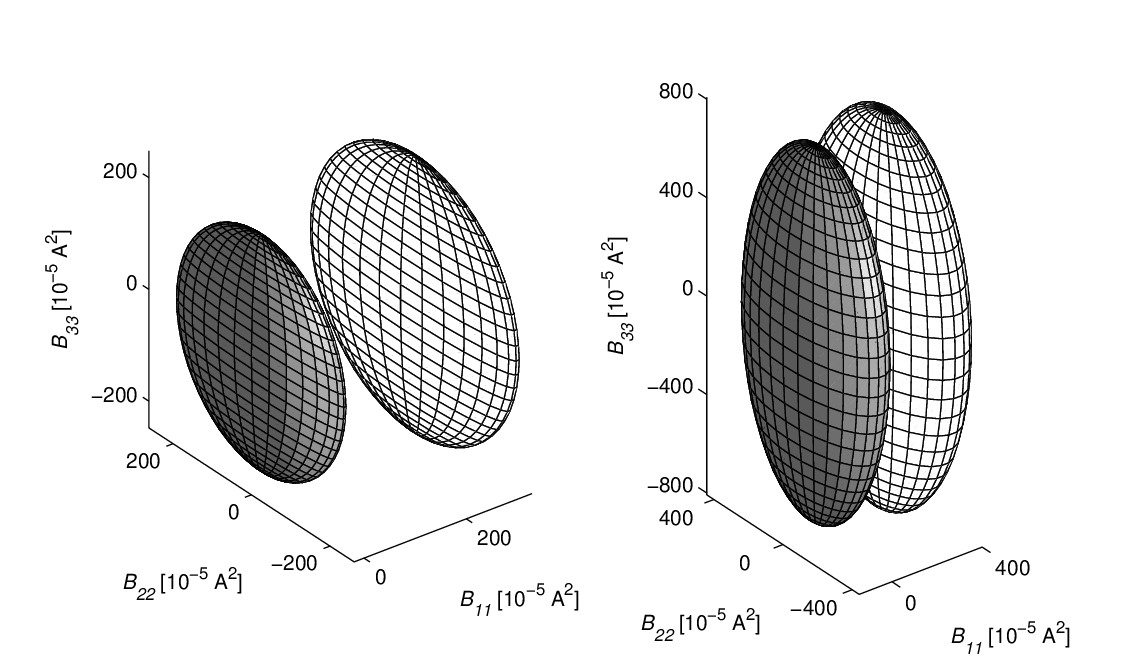}
 \caption{{Anisotropic thermal parameter for Cu in the $^{16}$O (gray) and  $^{18}$O (white) sample at 15~K (left panel) and 290~K (right panel).}
\label{ellipsoid}}
\end{figure}

\section{Discussion}

Our neutron diffraction study showed that the  samples La$_2$Cu$^{16}$O$_4$ and La$_2$Cu$^{18}$O$_4$ crystallized both in the orthorhombic space group {\it Bmba}, which is well known in literature~\cite{radaelli}. We would like to emphasize that they are single-phased which  confirms our results of the analysis of the oxygen content by means of hydrogen reduction which yielded an oxygen content of $4.004\pm0.005$. The small excess oxygen concentration explains the low values of the peak in the susceptiblity which is related to the N\'eel temperature $T_N$~\cite{kastner}.  The oxygen isotope substitution  raises the peak in the susceptibility  by $2.4\pm0.2$~K as opposed to results reported in Ref.~\cite{zhao}. However, we would like to point out that Zhao {\it et al.}~\cite{zhao} found that the values of $T_N$ and also the OIE on $T_N$ depends strongly on the preparation and annealing conditions of the samples.  
\\ 
The structural parameters obtained from our neutron diffraction measurements on La$_2$Cu$^{16}$O$_4$ are in excellent agreement with literature~\cite{jorgensen,radaelli}. Measurements on HRPT which is very well suited to determine lattice constants with highest accuracy revealed a negative OIE on the lattice constants, see Table~\ref{lattice_oie}. 
  We found a tendency for the Cu-O1 bond lengths to be larger in the $^{16}$O sample as compared to the $^{18}$O sample by the order of $10^{-4}$, see Table~\ref{lattice_oie}. No OIE could be detected regarding the atomic positions within the instrumental resolution and the accuracy of the Rietvield refinement.
\\
The values of the isotropic and anisotropic thermal parameters $B_{iso}$ and $U_{ii}$, $1\leq i \leq 3$ are in agreement with results reported in Refs.~\cite{radaelli,jorgensen} and in Ref.~\cite{chaillout}, respectively. 
We found a significant OIE on the zero point motion in terms of the thermal parameters up to approximately 50 \% at 15~K. The OIE still exists at room temperature, however it is substantially lower, its maximum is $\sim10$ \%,  see Table~\ref{biso_oie}. Interestingly, we observed a negative OIE on the thermal parameters - both $B_{iso}$ and $U_{ii}$, $1\leq i\leq 3$ - for O1 and O2, whereas the OIE is positive in case of La and Cu which is qualitatively consistent with the sum rule for lattice vibrations \cite{rosenstock}. Furthermore we would like to emphasize that the sign of the OIE on the anisotropic thermal parameters $U_{ii}$ is identical for the $i=1,2,3$ direction with the sign of the OIE on the $B_{iso}$ parameters for each atom. 
\\
Interestingly the OIE on the anisotropic thermal parameters varies significantly for the different direction of thermal motion at 15 K: $\Delta U_{11}\approx 0.9$ down to $\Delta U_{22}\approx 0.3$ and  $\Delta U_{11} \approx 0.5$ down to $\Delta U_{33}\approx 0.2$, for La and Cu respectively. In the case of the oxygen atoms O1 and  O2 the order of magnitude of the OIE on the anisotropic thermal parameter stays in a comparable range $\sim 0.2$. At 290 K the changes in the OIE are substantial for La and Cu: while the OIE on $U_{11}$ ($U_{33}$) vanishes within the present accuracy, we obtained $\Delta U_{33}\approx 0.2$ ($\Delta U_{11}\approx 0.3$) for La (Cu). The direction where the OIE is most pronounced is not identical with the direction where $U_{ii}$ is largest. 
\\
In order to visualize the effect of the isotropic thermal parameters we show in Figure~\ref{lattice_biso} the temperature evolution of the lattice parameter together with the mean displacement $a\pm a_0(Cu)$ deduced from the isotropic thermal parameters for the copper atom $B_{iso}(Cu)$ 
\begin{eqnarray}
a_0=\sqrt{\frac{B_{iso}}{8\pi^2/3}}. 
\end{eqnarray}
In this scale it is striking that  the OIE on the lattice parameter $a$ is not visible anymore, even the temperature dependence of $a\pm a_0$ as a whole expression is small. But the OIE on $a\pm a_0$ due to the movement of the copper ion can be observed in terms of $a$ being subject to a larger variation due to the thermal motion of Cu in the $^{18}$O compound than in the $^{16}$O compound.
\\
Figure~\ref{ellipsoid} illustrates the anisotropic thermal parameters for Cu in  the shape  of an ellipsoid with axes $U_{11}$, $U_{22}$ and $U_{33}$ corresponding to thermal motion. At 15~K the zero-point motion is almost isotropic in the $y$-$z$ direction, while there is a considerable anisotropic deformation at room temperature.  For O1 which is also in-plane the anisotropy is similar, the elongation in the $z$-direction is also most pronounced at 290~K while the zero point motion is almost isotropic. The out-of plane ions O2 and La behave in the different way: at 15~K La is almost isotropic in the $x$-$z$ direction and O2 is almost isotropic in the $y$-$z$ direction, whereas the maximum displacement is along the  $y$- and $z$-direction for La and O2, respectively. At room temperature the thermal motion is highly anisotropic and the largest elongation is along the $x$-direction for both out-of-plane ions La and O2.

\section{Conclusions}
In the present work we have studied the influence of the oxygen isotope effect ( $^{18}$O vs. $^{16}$O 78\%) on structural and thermal parameters by means of high-resolution neutron diffraction experiments on the antiferromagnetic insulator La$_2$CuO$_{4}$. The oxygen concentration in the $^{16}$O and $^{18}$O samples was determined to be $4.004\pm0.005$ by hydrogen reduction. The excess oxygen content was confirmed to be small by neutron diffraction measurements which revealed the samples to be single-phased ({\it Cmca}). We found a nonvanishing negative OIE on the lattice parameters in the order of 0.01\%. 
 The OIE on the Cu-O1 bond length turned out to be negative as well. We found a considerable OIE on the zero point motion in terms of the isotropic and anisotropic thermal parameters  (up to 100\% for La),
  it is positive for the atoms La and Cu, whereas it is negative for O1 and O2. A refinement using anisotropic thermal parameters yielded qualitatively the same result, in particular the sign of the OIE  on the anisotropic thermal parameters is identical for the $x$-, $y$- and $z$-direction with the sign of the OIE on the isotropic thermal parameters. Moreover, we discovered that the OIE on the anisotropic thermal parameters varies significantly with the direction of thermal motion e.g. $\Delta U_{11}\approx 0.9$ and $\Delta U_{11}\approx 0.3$ for La at 15 K. The huge OIE on the isotropic thermal parameters can therefore be attributed mainly to the thermal motion in the $x$-direction.
\\
Furthermore our diffraction study revealed that the anisotropic deformation of the ellipsoid of thermal motion is different for in-plane and out-of-plane ions. The elongation is most pronounced along the $z$-direction for Cu and O1 (in-plane) and along the $x$-direction for La and O2 at 290 K. Zero point motion is isotropic at least in two directions. The preferred direction of thermal motion is identical for the $^{16}$O and $^{18}$O sample.
\\
Thermal motion in La$_2$CuO$_4$ is in the range of 10-20 meV~\cite{boeni} whereas the energy scale of the antiferromagnetic exchange $J$ is of the order of 100 meV~\cite{coldea}. Thus the displacement of the ions due to thermal motion is almost static as compared to the electronic scale of the hopping integral $t$ and the antiferromagnetic exchange $J$. 
\\ \\
We hope that the present detailed structural data will stimulate theoretical studies on how the OIE on structural parameters as well as on the zero point motion affects the hopping integral $t$ and the antiferromagnetic exchange $J$ in La$_2$CuO$_4$.

\section{Acknowledgements}
We greatly acknowledge fruitful discussions with H.~Keller and B.~Batlogg as well as helpful contributions from D.~Hennig, B.~R\"ossner and U.~Blaesi. The study was supported by NCCR MaNEP Project.


\end{document}